\documentclass[traditabstract]{aa}

\newcommand{\xmm}{{\it XMM-Newton}}
\newcommand{\srg}{{\it SRG}}
\newcommand{\art}{ART-XC}
\newcommand{\ero}{{eRosita}}
\newcommand{\sxp}{{SXP\,1323}}

\newcommand {\be}{\begin {equation}}
\newcommand {\ee}{\end {equation}}

\usepackage{graphicx}
\usepackage{txfonts}
\usepackage{hyperref}
\usepackage{xcolor}
\usepackage{ulem}
\usepackage{academicons}
\usepackage{natbib}
\bibpunct{(}{)}{;}{a}{}{,} 

\newcommand{\orcid}[1]{\href{https://orcid.org/#1}{\textcolor[HTML]{A6CE39}{\aiOrcid}}}

\begin{document}

\title{Losing a minute every two years: {\it SRG} X-ray view on the rapidly accelerating X-ray pulsar \sxp}

   \author{I.A. Mereminskiy\thanks{i.a.mereminskiy@gmail.com}\inst{1} \and A.A. Mushtukov\inst{1,2} \and A.A. Lutovinov\inst{1} \and {S.S. Tsygankov}\inst{1,3} \and A.N. Semena\inst{1} \and S.V. Molkov\inst{1} \and A.E. Shtykovsky\inst{1}}
\titlerunning{\srg\, view on SXP1323}
\authorrunning{I. Mereminskiy et al.}
   \institute{Space Research Institute, Russian Academy of Sciences, Profsoyuznaya 84/32, 117997 Moscow, Russia
    \and
    Leiden Observatory, Leiden University, NL-2300RA, Leiden, The Netherland
    \and
    Department of Physics and Astronomy, FI-20014 University of Turku, Finland}

   \date{Received July 2021; accepted XXX}

 
\abstract{SXP 1323 is a peculiar high-mass X-ray binary located in the Small Magellanic Cloud, renowned for its rapid spin-up. We investigate for the first time broadband X-ray properties of \sxp\, as observed by the {\it Mikhail Pavlinsky} ART-XC and \ero\ telescopes on board the {\it SRG} observatory. Using \art\, and \ero\, data we produced first broadband 1-20 keV X-ray spectrum and estimated pulsed fraction above 8 keV.  With the addition of archival {\it XMM-Newton} observations we traced evolution of the SXP 1323 spin period over the last five years and found that after 2016 the source switched to a linear spin-up with rate of -29.9 s yr$^{-1}$. Broadband X-ray spectrum is typical for accreting X-ray pulsars, with steep powerlaw index ($\Gamma$=-0.15) and exponential cutoff energy of 5.1 keV. No significant difference between spectra obtained in states with and without pulsations were found.}

   \keywords{accretion, accretion disks -- magnetic fields -- pulsars: individual: \sxp -- stars: neutron -- X-rays: binaries}

   \maketitle

\section{Introduction}

The Small Magellanic Cloud (SMC) is a nearby irregular galaxy \citep[located at 62 kpc;][]{scowcroft16A}, which have undergo a major star-formation event about 50 Myr ago \citep{shtykovskiy07}. It resulted in the birth of a large population of high-mass X-ray binaries \citep[HMXBs, see, e.g. ][and Haberl et al. 2021, in preparation]{haberl16}, that are intensively observed over the last decades with various X-ray instruments \citep[e.g.][ and others]{cowley97,laycock10,haberl12smc}, including modern X-ray telescopes {\it Mikhail Pavlinsky} ART-XC \citep{art21} and {\it eRosita} \citep{2021A&A...647A...1P} on board the \srg\ mission \citep{sunyaev21}.

During the performance and verification phase \ero\, observed the supernova remnant (SNR) 1E0102-72.2, which is commonly used as a calibration target in soft X-rays. These observations were performed as a sequence of pointings, during several of them the bright X-ray pulsar \sxp\ was serendipitously observed by \art.

In this paper we report on the first investigation of \sxp\, broadband spectrum and its variability using \art\, and \ero\, data. We also employed archival {\it XMM-Newton} observations to trace the pulse period change over the last five years.

\section{SXP\,1323}

The \sxp\ was first detected in X-rays by the {\it ROSAT} observatory \citep{sasaki00, haberl00cat} and was proposed as an X-ray binary system with the Be companion (BeXRB) based on the positional coincidence with the bright $H_{\alpha}$ emission-line star \citep{haberl00ha}. Later, \cite{haberl05} discovered coherent pulsations from the source with a period of $\approx$1323 s, making this system one of the slowest BeXRB known. Using {\it Suzaku}, \xmm\, and {\it Chandra} \cite{carpano17} found a 26.2-days variability, which they identified as an orbital period. It is interesting to note that this period is unusually short for such systems -- from the Corbet relation \citep{corbet86} one could expect an orbital period around 300 days. The orbital modulation is clearly seen in X-rays, leading to the variations of the source luminosity from $10^{35}$ to $few \times 10^{36}$ erg s$^{-1}$. An equivalent width of the $H_{\rm \alpha}$ line of the optical companion of \sxp\ follows the known $P_{\rm orb}-H_{\rm \alpha}$ correlation \citep{reig97}, thus confirming the identification of 26.2 day period as an orbital one \citep{gvaramadze19}.

\citet{carpano17} noticed a peculiar evolution of the neutron star (NS) spin period, which was relatively stable during 2003-2006 and then started to spin-up very rapidly, with steady $\dot P \simeq -21.65$ s yr$^{-1}$. This makes \sxp\, one of three pulsars with largest ever observed spin-up rates, along with the peculiar SXP\,1062 \citep{henault12,tsygankov20sxp1062} and NGC300 ULX-1 \citep{carpano18ngc300,vasilopoulos19}. A similar behaviour in the form of a transition from the plateau to the rapid spin-up was also observed in the long-period BeXRB 2RXP\,J130159.6-635806 \citep{krivonos15}.

In archival $H_{\rm \alpha}$ images \citet{gvaramadze19} found a bright circular shell which they proposed as a supernova remnant (SNR), that was created during the formation of \sxp. Later, the SNR nature of the shell was confirmed based on the deep radio and X-ray imaging \citep{maggi19}. If this association is correct, it allows to estimate the age of \sxp\, to $25-40\times10^{3}$ years, making it one of the youngest known BeXRBs with the visible supernova remnant \citep[see][for a list of know XRB with associated SNR]{maitra21}. 

\section{X-ray observations}

We analysed the \art\, observation of 1E\,0102.2-7219 (ObsID: 700001001000), started on Nov 7, 2019 and lasted for $\sim60$ ks. According to the orbital solution obtained by \citet{carpano17} our observation happened at the orbital phase of $\approx0.7$, corresponding to the maximal X-ray brightness of the system. In this state the observed 0.3-10 keV luminosity reaches $1-2\times10^{36}$ erg s$^{-1}$ and the source spectrum becomes harder, therefore providing optimal conditions for the \art\, observations. 

\art\ data were processed with the standard pipeline {\sc artproducts} v0.9 with the {\sc caldb} version 20200401. Although \art\, works in the broad 4-30 keV energy band, for spectral analysis we used only 5-22 keV  range, given the still present calibration uncertainties at low ($<5$ keV) energies and lack of a significant source detection above 22 keV. We have also produced the source lightcurve in the 4-12 keV energy band for timing analysis. 

The SNR 1E\,0102.2-7219 itself was not detected in the \art\, data above 5 keV. Nevertheless we have explicitly checked, that the SNR is not contributing to the \sxp\, spectrum. First, the nearest part of the shell is located $\approx1.5\arcmin$ away from the position of \sxp\, and therefore the \art\, point-spread function is enough to spatially resolve them, limiting a possible contribution from SNR. Next, we estimated the hard X-ray flux from SNR using a standard spectral model for this object \citep{plucinsky17}. Its overall flux is about $F_{\rm 5-10~keV}\approx 8 \times 10^{-14}$ erg cm$^{-2}$ s$^{-1}$, which is more than an order of magnitude lower than the flux registered from \sxp. 

One of the most interesting properties of this pulsar is its spin evolution. As it was shown by \cite{carpano17} after 2006 \sxp\, begun to spin-up at impressive rate of $\dot P \simeq -21.65$ s yr$^{-1}$, and kept accelerating until 2017. In order to trace this behaviour after 2017 we collected few {\it XMM-Newton} observations performed in 2016-2020 (ObsIDs: 0412983201,0412983401, 0810880101, 0810880601). Using {\it XMM SAS v19} and corresponding calibration files we extracted barycentered source lightcurves in the 1-10 keV energy band, with excluded periods of an enhanced particle background. 

In order to extend \art\, spectrum to a softer X-ray band we have extracted the source spectrum from the simultaneous \ero\,  observation using {\it eSASS} package \citep{brunner2021erosita}. Given the possible contribution from 1E\,0102.2-7219 at low energies and presence of the variable soft component \citep{carpano17}, we  limited the \ero\, data to the 1-7 keV range. Both source spectrum and background were extracted from regions with $25\arcsec$ radius, located at same distance from the center of SNR 1E\,0102.2-7219. Spectra from seven telescope modules were joined together and grouped in order to have at least 30 counts per energy bin.

We also extracted the spectrum from the {\it XMM-Newton} EPIC-pn observation 0412981001, which was carried out in 2010, using the aperture with $r=10\arcsec$ in the energy band 1-10 keV. Note, that we did not tried to use EPIC-pn data below 1 keV for similar reasons as with \ero. The spectrum was grouped in order to have at least 30 counts per energy bin. To fit the source spectra the $\chi^2$ statistics was applied.

\section{Results}

\subsection{X-ray spectrum}

\art\, is the first  grazing incidence telescope ever to observe \sxp\, at energies above 10 keV, providing the possibility to better characterise the source broadband spectrum.  

Following the spectral analysis by \citet{carpano17} we tried to fit a joint \art + \ero\, spectrum in the 1-22 keV energy band with a simple {\texttt const*phabs*vphabs*powerlaw} model in the {\sc XSPEC} package. The multiplicative constant allows to take into account the difference in the calibrations between \ero\, and \art. The {\texttt phabs} component, describing the Galactic absorption, was fixed at the value of $5.36\times10^{20}$ cm$^{-2}$ \citep{dickey90}, while the \texttt{vphabs} component, responsible for the local SMC absorption, was allowed to be variable with the fixed abundances of all elements (A$_{\rm He}=1$, A$_{Z>2}=0.2$, \citealt{russell92}). This model obviously over predicts the hard X-ray flux above $\approx10$ keV. Therefore, we added an exponential cutoff at higher energies ({\texttt const*phabs*vphabs*cutoffpl}) as usually observed in X-ray pulsars \citep[see, e.g.][]{coburn02,filippova05}. 

\begin{figure}
    \centering
    \includegraphics[width=\columnwidth]{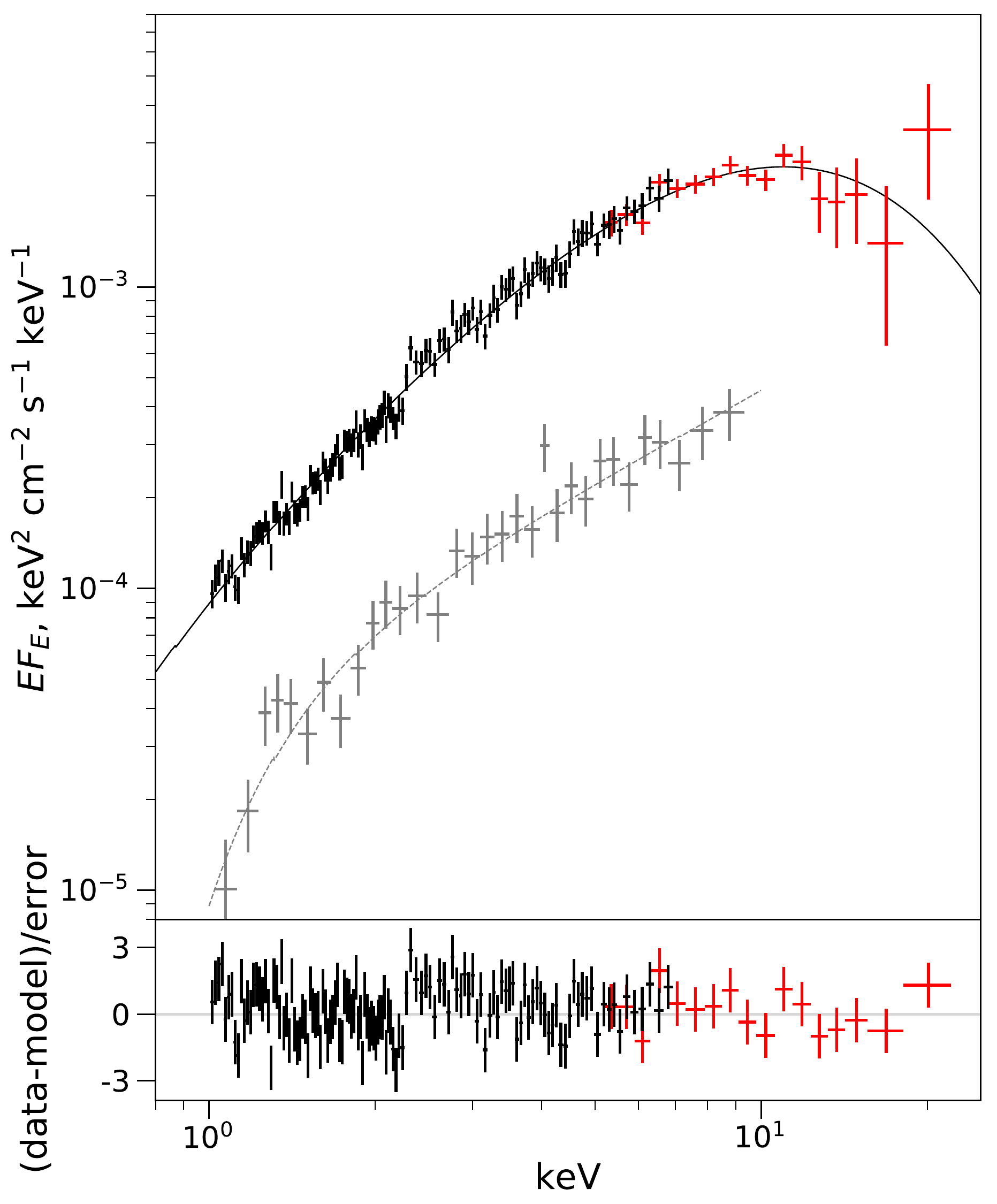}
    \caption{Joint \art\ + \ero\, spectrum of \sxp. Black points correspond to the \ero\, data, red points are \art\, measurements. Spectra from seven \art\, modules were grouped together and rebinned for clarity. Gray points show the {\it XMM-Newton} source spectrum, obtained in 2010, when no pulsations were detected.}
    \label{fig:spectrum}
\end{figure}

As it could be seen from Fig.~\ref{fig:spectrum}, the resulting model provides a good approximation to the \sxp\ spectrum with $\chi^{2}/{\rm d.o.f.} = 498.4/444$. The model was insensitive to the local SMC absorption, with only an upper limit of $5\times10^{20}$ cm$^{-2}$, we therefore fix it at $10^{20}$ cm$^{-2}$. The photon index was measured to be $\Gamma = -0.15^{+0.07}_{-0.07}$, and the cutoff energy to be $E_{\rm cut} =5.1^{+0.7}_{-0.5}$ keV. No obvious features that could be ascribed to a cyclotron absorption line are obviously seen. The total unabsorbed bolometric luminosity, estimated over the 0.1-100 keV energy band, is $4\times10^{36}$ erg s$^{-1}$.

Another peculiar trait of \sxp\, is a disappearance of the pulsations in 2010-2014 which was noticed by \citet{carpano17}. It was proposed by \cite{clk18} that the neutron star magnetic field in \sxp\, is relatively weak, below $10^{11}$ G, and was crushed by the infalling material at some moment around 2005. It is interesting therefore to compare the X-ray spectrum of \sxp\, observed during the lack-of-pulsation phase with the observed joint \art\ + \ero\, spectrum. If the accretion disk reaches all the way down to the NS surface it should form some sort of a boundary (transition) layer \citep{inogamov99}, as observed in low-mass X-ray binaries at high accretion rates \citep[see, e.g.][]{revnivtsev13}. This boundary layer usually manifests itself in the source X-ray spectra as a thermal component with a temperature of $kT\approx2.5$ keV.
We have used the {\it XMM-Newton} observation 0412981001, obtained at the orbital phase of 0.54, close to the phase of the \srg\, observation, to check for the presence of such a thermal component in the \sxp\ spectrum. The resulting spectrum is well described with a similar {\texttt phabs*vphabs*powerlaw} model in the 1-10 keV band, without need for any additional components above 1 keV. The spectrum is hard ($\Gamma\approx1$) and the absorption column thickness is significantly higher ($N_{\rm H}\approx 2\times10^{22}$ cm$^{-2}$) as expected from the observed orbital modulation, close to the parameters derived by \citet{carpano17}. We could therefore conclude, that the spectral shape during the non-pulsation episode was not dramatically different. 

\subsection{Pulsations -- spin-up rate and pulsed fraction}

In order to trace the spin-up history of \sxp\, in 2016-2020 we have searched for periods in the \art\, and {\it XMM-Newton} lightcurves using a simple epoch folding technique \citep{leahy83}. Pulsations were clearly detected in all observations with the period ranging from 1094 to 976 seconds, correspondingly. In order to estimate an uncertainty for the period determination we performed a large number of simulations for each lightcurve using the observed count statistics 
\citep[see details in][]{boldin13}.
We also accounted for the uncertainty caused by the orbital motion of the neutron star and found that in the worst-case scenario (high inclination, large eccentricity) it could be about 1 s. 

\begin{figure}
    \centering
    \includegraphics[width=\columnwidth]{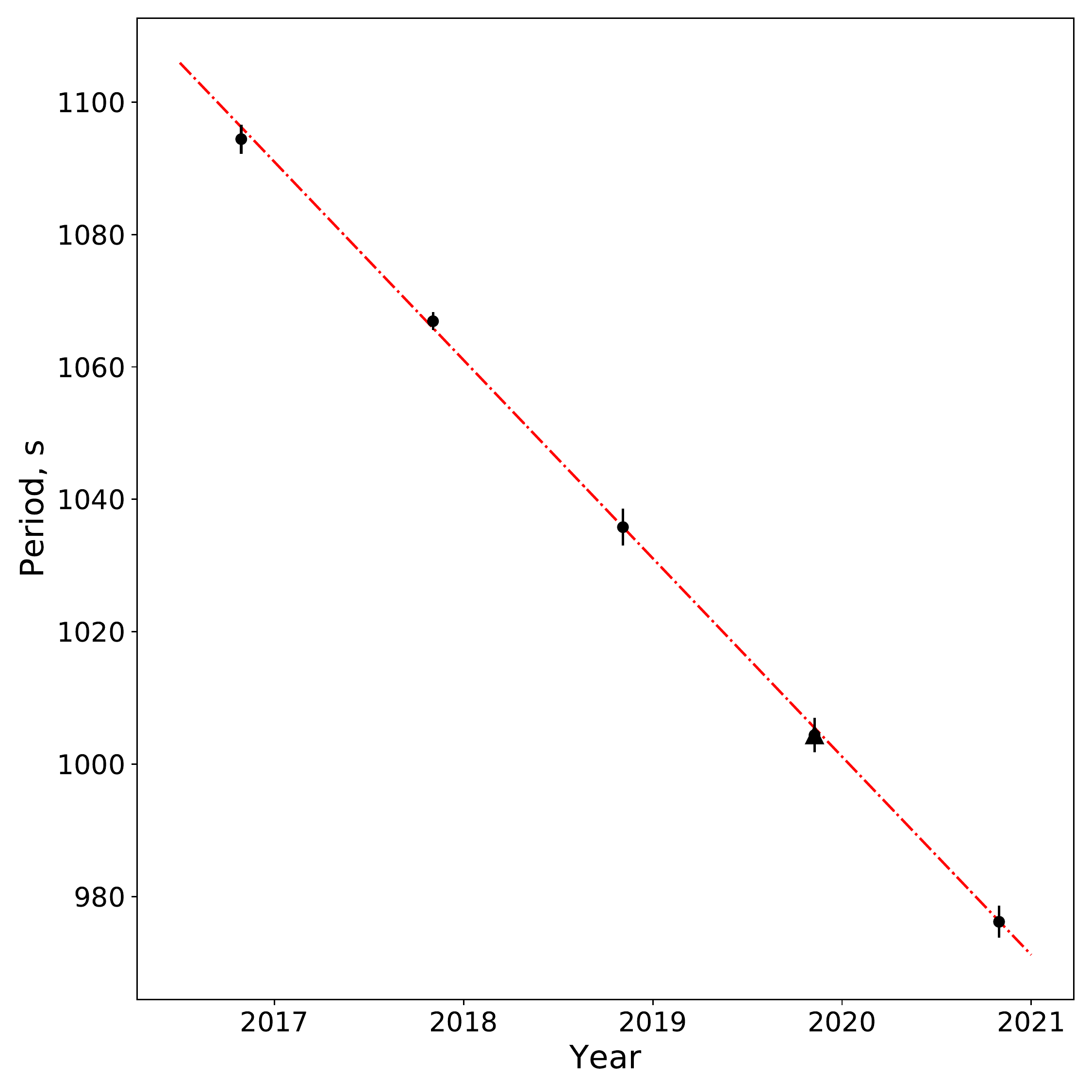}
    \caption{Long-term period evolution of \sxp. Circles show {\it XMM-Newton} measurements and triangle denotes \art\, data. Dash-dotted line shows the best-fit approximation with the spin-up of -29.9 s yr$^{-1}$.}
    \label{fig:perevol}
\end{figure}

Fig.~\ref{fig:perevol} shows the long-term evolution of the spin period along with the best-fit linear model with the spin-up rate of $\dot P =  -29.9\pm0.5$ s yr$^{-1}$.  This value is significantly larger than one reported by \cite{carpano17} over the 2006-2016 period ( $\dot P =  -21.65$ s yr$^{-1}$), that could indicate that at some moment after 2014 the source switched to the linear and faster spin-up.

\begin{figure*}
    \centering
    \includegraphics[width=17cm]{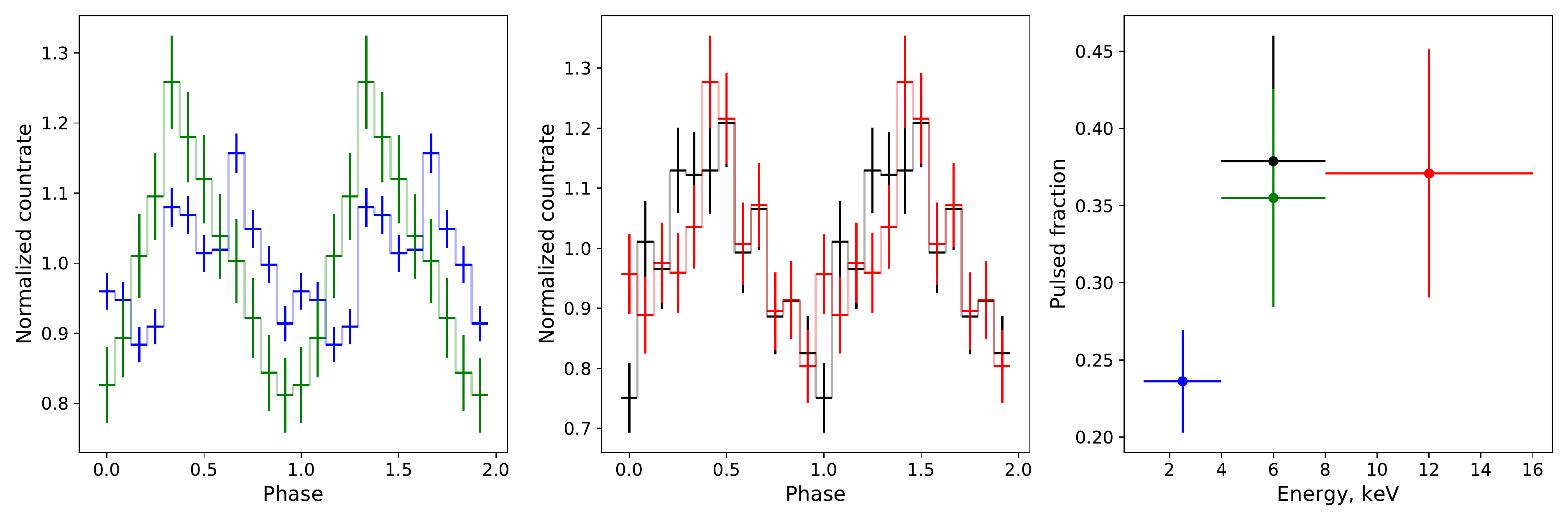}
    \caption{Left panel: a pulse profile in the 1-4 keV (blue) and 4-8 keV (green) band as seen by \ero. Middle panel: a pulse profile in the 4-8 keV (black) and 8-16 keV (black) band as observed by \art. Right panel: measured pulsed fractions $PF$.}
    \label{fig:pulse}
\end{figure*}

It is also interesting to look at the source pulse profile at different energies. On Fig.~\ref{fig:pulse} we present pulse profiles in three broad energy bands as seen by \ero\, and \art. The \ero\, pulse profiles are shown in the 1-4 and 4-8 keV energy bands and \art\, profiles are in the 4-8 and 8-16 keV bands. In the lowest energy band the pulse profile demonstrates complex shape, while at higher energies it is simpler, with characteristic single peaked structure. At the highest energies the maximum becomes narrower, with less prominent right shoulder \citep[see also Fig.~1 in ][]{yang18pf}.

We calculated the pulsed fraction as specified in \citet{yang18pf}: 
\begin{equation}
PF_{\rm A} = \frac{f_{\rm max}-f_{\rm min}}{f_{\rm max}}
\end{equation}

Dependence of the pulsed fraction on energy shows clear increase from $\sim25$\% in the 1-4 keV energy band to $\sim38$\% in the 4-8 keV band 
and a similar value at higher (8-16 keV) energies. The observed $PF_{\rm A, 4-8 keV}\approx0.38$ is close to $PF_{\rm A}$ measured in the archival {\it Chandra} and {\it XMM-Newton} data by \citet{yang18pf} at similar orbital phases.

Alternatively, we estimated also the pulsed fractions in the standard way as 
\begin{equation}
PF = \frac{f_{\rm max}-f_{\rm min}}{f_{\rm max}+ f_{\rm min}}
\end{equation}

Obtained values $PF_{\rm 1-4 keV}\approx13$\%, $PF_{\rm 4-8 keV}\approx PF_{\rm 8-16 keV}\approx23$\% are slightly lower than  mentioned above, but demonstrate the same tendency.

\section{Discussion}

\srg\, observations provide the first ever high quality broadband X-ray spectrum of \sxp\, which turned out to be quite usual for accreting neutron stars in high-mass X-ray binaries. Much less usual is the spin-up history of this peculiar source. After the plateau in early 2000s the source started to spin-up very rapidly. According to our measurements over last five years the source accelerates with even higher rate of $\dot P \simeq -30$ s yr$^{-1}$. Such a high acceleration rate of the rotation of the neutron star at a rather low luminosity allows to make theoretical assumptions and estimates of the physical processes of the accretion in this system. 

First of all long-term observations of the spin period derivative available up to date in \sxp\ make it possible to estimate the dipole component of NS magnetic field.
In the case of accretion through the disc, the spin period derivative is determined by the mass accretion rate $\dot{M}$, current spin period $P$, NS moment of inertia $I$ and magnetospheric radius $R_{\rm m}$ (see, e.g., \citealt{1972A&A....21....1P}):
\be\label{eq:Pdot} 
\dot{P}=\frac{\dot{M}}{I}\left[
2 P R_{\rm m}^2 - \frac{P^2}{\pi}\sqrt{GMR_{\rm m}}
\right].
\ee 
The magnetospheric radius is dependent on the NS magnetic field strength and accretion luminosity: 
\be 
R_{\rm m}=2.4\times 10^{8}\,\Lambda\, B_{12}^{4/7}L_{37}^{-2/7}m^{1/7}R_6^{10/7} \,\,\,{\rm cm}, 
\ee 
where $B_{12}$ is the magnetic field strength $B$ at the NS poles in units of $10^{12}\,{\rm G}$,
$L_{37}$ is the accretion luminosity $L$ in units of $10^{37}\,{\rm erg\,s^{-1}}$,
$m$ is the mass of a NS $M$ in units of masses of Sun,
$R_6$ is the NS radius $R$ in units of $10^6\,{\rm cm}$, and
$\Lambda$ is a constant dependent on accretion flow geometry: $\Lambda=1$ for the spherical accretion and $\Lambda<1$ for the accretion from the disc with $\Lambda=0.5$ being a commonly used value (see e.g., \citealt{2002apa..book.....F}).  
If NS spin period is close to the equilibrium, the corotational radius $R_{\rm c}$ is close to the magnetospheric radius $R_{\rm m}$, and terms in the right-hand side ({\it rhs}) of Eq.\ref{eq:Pdot} compensate each other. 
If magnetospheric radius locates deep inside the NS magnetosphere, one can neglect the first term in {\it rhs} of (\ref{eq:Pdot}).
As a result, we get an estimation for the surface B-field strength:
\be 
B_{12}\approx \left(\frac{\dot{P}}{5.7\times 10^{-6}\,{\rm s\,s^{-1}}}\right)^{7/2}\,
\Lambda^{-7/4}\dot{M}_{17}^{-3}P_3^{-7}
I_{45}^{7/2}m^{-3/2}R_6^{-3}
\ee 
Using the observed spin period and spin period derivative, and estimating the average mass accretion rate as $\dot{M}_{\rm ave}\approx 10^{16}\,{\rm g\,s^{-1}}$ we get a rough estimation of the surface magnetic field strength: $B\sim few\times 10^{13}\,{\rm G}$. 

The physics of torques, however, can be much more complicated (see, e.g., \citealt{2014EPJWC..6401001L}).
According to the model by \citealt{1978ApJ...223L..83G}, the net total torque on the NS can be written as
\be 
K_{\rm s}\approx n(\omega_{\rm s})\dot{M}\sqrt{GMR_{\rm m}},
\ee 
where 
\be 
n(\omega_s)\simeq 1.4\left(\frac{1-\omega_{\rm s}/\omega_{\rm c}}{1-\omega_{\rm s}}\right),
\ee
the fastness parameter $\omega_{\rm s}=(R_{\rm m}/R_{\rm c})^{3/2}$, and the critical frequency $\omega_{\rm c}\approx0.8$ \citep{1988ApJ...328L..13L}.
Direct application of this model to the data give as $B\approx 6\times 10^{13}\,{\rm G}$.

Note that both estimations are rough and strongly dependent on the exact geometry of the accretion flow and displacement of the accretion disc inner radius (this uncertainty is involved in the parameter $\Lambda$). 
Nevertheless, the obtained estimation of the field strength allows us to conclude that the source is always in the sub-critical regime of accretion when the radiation pressure is not enough to affect the accretion flow significantly above the NS surface \citep{1976MNRAS.175..395B,2015MNRAS.447.1847M}. 
The X-ray pulsar with the surface field strength $>10^{13}\,{\rm G}$ and spin period of $\sim 1000\,{\rm s}$ cannot turn in to the ``propeller" state if it accretes from the disc. 
Instead of that, it turns into the state of a stable accretion from the cold disc composed of the recombined material  \citep[][]{2017A&A...608A..17T}.

However, the physical picture in \sxp\, can be more complicated due to the short orbital period and relatively small separation between components in the binary system.
According to numerical simulations \citep{2014ApJ...790L..34M}, the decretion disc around the Be-companion can be disturbed by the neutron star. In this case, both the decretion disc around Be-star and the accretion disc around the NS are not necessarily aligned with the orbital plane, and one can speculate that the orientation of the disc can vary in time. 
Note that the equation (\ref{eq:Pdot}) is written under the assumption of the aligned spin axis of the accretion disc and NS. 
If there is a non-zero angle between two axes, the accretion process affects mainly the component of the NS angular momentum aligned with the axis of the accretion disc. 
Then one can observe the spin period behaviour reported by \citealt{carpano17}, when the NS spin-down switches to the NS spin-up with the small but non-zero spin frequency change between these two stages. 
The minimal spin frequency variations observed by  \citealt{carpano17} in this case correspond to the situation when the rotational axis of the neutron star belongs to the accretion disc plane. 
Variations of the NS spin axis direction can, in principle, cause variations in the pulsed fraction and, as a result, the disappearance of pulsations detected in 2010-2014 by \citealt{carpano17} (this statement, however, requires numerical confirmations).

Taking into account a short orbital period of \sxp\ and the nature of the companion star, we cannot exclude a contribution of a collimated (by the NS orbital motion) wind accretion onto the NS surface (see, e.g., \citealt{2019A&A...622L...3E}).
Accounting for the wind accretion changes the estimation of the magnetospheric radius and slightly changes the estimations for the B-field strength.

The uncertainties in magnetic field estimations can be clarified by further observations of spin period evolution, where one would expect the appearance of the measurable second spin period derivative. 
Additional information one can get from the investigation of X-ray spectra at low-luminosity state at energies $>10\,{\rm keV}$. 
Recently, it has been shown that the spectra of XRPs with extremely strong surface magnetic fields experience dramatic changes under conditions of extremely low mass accretion rates \citep{2019MNRAS.483L.144T,2019MNRAS.487L..30T,2021MNRAS.503.5193M,2021ApJ...912...17L,2021A&A...651A..12S}, and the analyses of the broadband X-ray spectra can help to limit the field strength.


\begin{acknowledgements}

This work is based on data from {\it Mikhail Pavlinsky} ART-XC and eROSITA, X-ray instruments on board the SRG observatory. The SRG observatory was built by Roskosmos in the interests of the Russian Academy of Sciences represented by its Space Research Institute (IKI) in the framework of the Russian Federal Space Program, with the participation of the Deutsches Zentrum für Luft- und Raumfahrt (DLR). The ART-XC team thank the Russian Space Agency, Russian Academy of Sciences and State Corporation Rosatom for the support of the \srg\ project and \art\ telescope and the Lavochkin Association (NPOL) with partners for the creation and operation of the \srg\ spacecraft (Navigator). The eROSITA X-ray telescope was built by a consortium of German Institutes led by MPE, and supported by DLR.  The science data are downlinked via the Deep Space Network Antennae in Bear Lakes, Ussurijsk, and Baykonur, funded by Roskosmos. The eROSITA data used in this work were processed using the eSASS software system developed by the German eROSITA consortium. This work was supported by the grant of the Ministry of Science and Higher Education of the Russian Federation 14.W03.31.0021.

\end{acknowledgements} 

\bibliographystyle{aa} 
\bibliography{biblio.bib} 
\end{document}